# An Initial Description of Capabilities and Constraints for a Computational Auditory System (an Artificial Ear) for Cognitive Architectures

**Frank E. Ritter**	FRANK@FRANKRITTER.COM
**Mathieu Brener**	MATBRENER@GMAIL.COM
Applied Cognitive Science LLC, State College, PA USA

**Jeffrey B. Bolkhovsky**	JEFFREY.B.BOLKHOVSKY.CIV@MAIL.MIL
Naval Submarine Medical Research Labrartory, Groton, CT USA

## Abstract

We present an initial set of factors, features, and constraints for developing a Computational Auditory System (CAS, aka less formally an artificial ear, AE) for use by cognitive architectures. We start to define a CAS and what tasks it should be able to perform. We then outline the features of a CAS for use by a cognitive architecture and factors that influence its performance. We conclude with an update on what has been created so far and insights on how to create and use a CAS in a cognitive architecture and include a set of functionalities for an artificial ear.

## 1. The Design of a Computational Auditory System

Cognitive architectures and AI agents have had difficulties interacting with the outside world. For example, there is work to provide these agents with artificial eyes and hands. A next step is to look at providing a form of hearing for these agents as well. We provide an initial design for a Computational Auditory System (CAS). An artificial ear would allow models in a cognitive architecture to perform a large new set of tasks that they have not been able to do before. For example, to hear things, to interact with experimental apparatus that puts out a beep for a new trial, to participate in auditory vigilance tasks, to work with auditory alarms (Ritter, Baxter, & Churchill, 2014, Ch. 4.6; Stanton, 1994), to be a more intelligent and grounded (Tehranchi, 2021b) agent in video games, to interact with other agents in the world (Trafton et al., 2006), and to recognize its name when spoken to. An artificial ear would thus help ground models and agents in the world.

    We start by quickly describing cognitive architectures because computational auditory systems will provide them with an artificial ear, in the same way that previous work provided them with artificial eyes and hands (e.g., Byrne, 2001; Ritter, Tehranchi, Dancy, & Kase, 2020; Tehranchi, 2021a). We then briefly review previous work on providing auditory input to cognitive architectures. Finally, we describe an initial design for the inputs and outputs for a computational auditory system, and an initial set of general constraints on a computational auditory system for

use by cognitive architectures.

## 2. Cognitive Architectures

Cognitive architectures can be seen as a set of fixed mechanisms used to model cognition. They provide a way to summarize and apply the results of empirical studies or a way to combine multiple smaller theories to see their joint implications (Byrne, 2003; Newell, 1990), and typically provide general agent capabilities. One can also see them as a type of constrained programming language (the symbolic ones) or as a set of neural networks (connectionist and neural ones). Many are a combination, called hybrid architectures (Sun, 1996). ACT-R is an example. Architectures can also be seen as a probabilistic programming language that can be noisy or errorful and that can improve with practice (Norling & Ritter, 2004).

We briefly note three types of architectures that might use the simulated ear.

ACT-R is a hybrid architecture that uses rules to represent procedural knowledge, and tuples to represent declarative knowledge (Anderson, 2007; Ritter, Tehranchi, & Oury, 2019). The rules match against declarative knowledge in a declarative memory store and in a goal stack. The system can learn by repeatedly accessing the declarative knowledge to strengthen it and make it faster. It can learn by rules firing close enough together to get merged into a larger, faster rule. ACT-R has a perceptual component (Byrne, 2001), but it does not include any hearing except that there is a location for an ear to put auditory information into.

Soar (Laird, 2012) is a more symbolic cognitive architecture, although it has moved towards being more hybrid. Models in Soar in the past that have interacted with a world have used perception and motor output modules in a somewhat ad hoc manner. An ear could be and has been defined to put auditory information into the top state.

Arcadia is an architecture that explores the role of attention in cognition (Bridewell, Wasylyshyn, & Bello, 2018; O'Neill, Bridewell, & Bello, 2018). Its knowledge structures are related to plans or strategies in BDI (belief-desires-intentions) architectures in that they represent plans more than single operators or rules. It appears to have finer-grained perceptual and cognitive knowledge constructs as well. We have started to integrate a preliminary implementation of a computational audio system to work with Arcadia (Gever et al., 2020). The artificial ear puts sound information, including object number (an ID), time of occurrence, frequency, azimuth, and duration, into an auditory short-term memory.

## 3. Prior Published Work on Modeling Ears and Audition

Sound has not often been used in cognitive models or architectures. For example, in Newell (1990) and Anderson (1983), "hear*", "aud*", and "sound" do not appear in the index. In Anderson (2007), the auditory system is mentioned but not used extensively. There is an auditory system in ACT-R/PM, but it is simply a buffer to put symbols into, not a wire to a microphone. Some reviews do not mention sound or auditory systems (Ritter et al., 2003) and others do (Pew & Mavor, 1998). Grossberg's neural network framework, ART, appears to be an exception, because it can perform some auditory scene analysis (Grossberg, Govindarajan, Wyse, & Cohen, 2004), and has lessons for how to implement an auditory scene analysis system.

Kotseruba and Tsotsos (2020) note in their extensive review several architectures that include auditory processing at a simulation level (e.g., ACT-R), several that include simulated physical sensors, and others that include physical sensors. The ACT-R, Soar, and EPIC architectures include only simulated sensors (i.e., a hearing module can place a token into memory), but do not include active listening. The systems that include physical sensors should be examined for guidance, but they are not architectures typically seen at conferences on cognitive modeling and architectures such as the International Conference on Cognitive Modeling (ICCM) or Behavioral Representation in Modeling and Simulation (BRIMS). The first several architectures in their figure of architectures and features that have audition (Glair, DAC, T3, Ymir), at least in some references to these architectures, do not hear from a physical sensor, but these appear to simply receive auditory inputs as symbols (GLAIR, DAC) or only recognize speech (T3, Ymir). Further review could be done here through the 17 noted as having audition.

To find capabilities and factors used by previous models we checked (as much as possible) the *CogSci* (since 1979), *ICCM* (since 1998), and *Artificial General Intelligence* (since 2008) proceedings, and Google Scholar for "cognitive architecture" and "cocktail party effect" or "hearing" or "ear". We were unable to find papers that directly discuss an artificial ear and cognition within the context of a cognitive model. There are papers on natural language understanding as text, though. These lack of publications for hearing can be contrasted with modeling the Stroop effect in vision, where there are thousands of papers on the effect in general and several papers on how it would appear in a cognitive architecture. In addition, the models of natural language understanding is only part of an auditory scene and only captures only one type of stimulus, akin to if you only captured text from a visual scene. A useful start, but not a unified model of hearing.

## 4. A functional ear

Figure 1 provides a basic block diagram for a computational audio system. This diagram presents a specification, not yet a created system. It also does not specify how such a system should be implemented (e.g., neural network). We are working on creating such a system.

The ear itself (on the left) interfaces with the world, either through a microphone or microphones, or via a wire to a sound source. The sound source or "ear" may have a pinna (the external ear, what wiggles when you wiggle your ears) to better mimic human hearing. It might be modifiable to model different types of ears. There will be some value to allowing this ear to use different inputs for development, testing, and deployment, such as sets of prerecorded audio files of varying complexity and a live microphone. The ear module will pass information into the preprocessor, which is likely to have extensive memory and processing capabilities. The preprocessor can keep state of what is being listened for (top-down or intrinsic) and do the basic match between sounds and cognitive constructs. This information can probably be passed internally to the ear system.

The preprocessor has to pass its outputs to cognition. These outputs may have to be modified by the processor to support the representations used by the cognitive architecture, and they will have to be passed to the architecture using a connection.

The connection between the processor and cognition will vary based on the architecture and the system. The ear can be a function call to the cognitive architecture when the ear and architecture are implemented in the same programming language, or it might be a socket or Unix pipe. JSON can be a useful way to connect systems as well (Hope, Schoelles, & Gray, 2014).

The preprocessor may have multiple processes in it because it has many parallel tasks to perform concurrently. Alexa and other voice recognition systems do not address this problem of attention shifts, as they do not have to process another stream concurrently and are only looking for their name. Running a separate process to look for one's name is possible, but probably leaves out a lot of features of general listening.

Cognition interacts with the preprocessor. While cognition will vary by architecture it can be expected to provide a way to have intrinsic expectations about what sounds will occur and when. Cognition can also hold memory systems that can provide expectations as well as a way to produce behavior with respect to what is heard.

This block structure can mirror the structure we have seen in simulated eyes and hands (Ritter, Baxter, Jones, & Young, 2000). Thus, there is also a role for a control panel or for a controller harness to specify what to listen to and what is heard, where the ear is demonstrated using ambient sounds or a file of sounds (played concurrently to the listener and the CAS). These tools can be useful when developing a system to test and exercise its parts more directly than running it as a complete system. More details on these types of testing and control systems are available in reports on using a simulated eye (Ritter, Van Rooy, & St. Amant, 2002).

Listening can thus be driven by both endogenous and exogenous attention. Endogenous attention describes what a listener selects top-down, or rather, in a goal- or expectation-driven manner from cognition. These could include factors related to following a given speaker or listening for a series of sounds tied to an expected task. Exogenous attention describes selection based on factors outside the listener's goals or expectations, and include sound types or levels that draw a listener's attention to the stimulus automatically and are passed to cognition asynchronously.

Figure 2 provides a functional schematic of a possible way to create an artificial ear for a cognitive architecture. Sekuler and Blake (2001) use this diagram to explain hearing. It starts with a hardware input, with binaural inputs. Their differences and their loudness and frequency are computed (the first blocks) and used by all later stages as well as being passed directly to perception within cognition. The loudness and frequency as well as interaural time and intensity are passed to a localization module. Localization is passed to perception as well as to the next stage of sound identification.

| Ear | ⇐==⇒ | Preprocessing | ⇐======⇒ | Cognition |
|---|---|---|---|---|
| Constraints on the mechanics of sound are applied. | Part of the CAS | Preprocessing and also the memory of targets and non-targets. Primarily matches sounds to targets from cognition. What to do with unknown sounds has to be an important area. | Passes commands and results between Pre-processor and Cognition. Depends on the base language of the cognitive architecture. | May pass information to the CAS. It will need to know what it can pass to the ear to listen to, and then use what is passed back. |

Fig. 1. Diagram of the functional components for a computational audio system. Arrows indicates connections.

Figure 2 suggests that an artificial ear, like a natural auditory system, might not be best represented or implemented as a monolithic system, but that it might have several stages where the sound information is processed. It is also the case that there may be further substructures that would be useful in creating a computational audio system.

The mechanisms in Figures 1 and 2 would allow the pre-processor to be loaded with a set of permanent words (including the agent's name) and perhaps a few to a few dozen words that would also be recognized (family members' names, reserved words from work), and the ability to load in context specific words.

We would anticipate that the loaded in words would decay with time because all aspects of memory decay. As they decay, the ability to recognize them would decline. That is, recognition would be slower and less likely. The decay rate could be dependent on their base activation in cognition, or there could be a different rate in auditory memory. This difference suggests that a study could be run to measure this type of auditory individual differences.

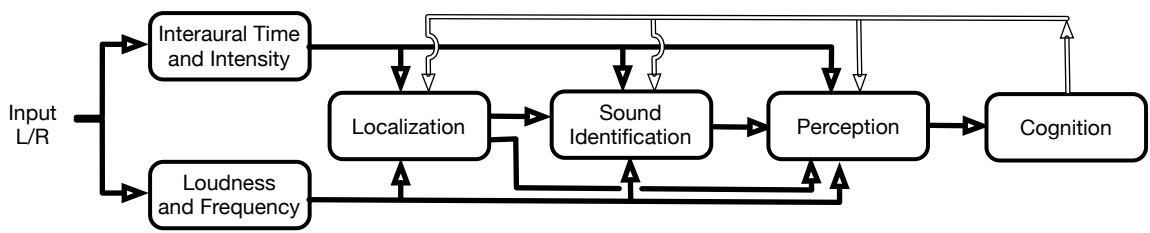

Fig. 2. Diagram of the potential substructure of functional components for a pre-processing component of a computational audio system, based on Sekuler and Blake (1985, Figure 9.8).

Table 1 notes some functions that a simulated ear might provide, essentially a high-level outline of how it will work. These would serve as function points in a programming language or operators in an architecture. Note, this set of functions references some internal state variables, such as features to listen for and features to not listen for, which would be part of the system as well.

*Table 1.* Suggested functions of a model human ear for use with a cognitive architecture and agents.

---

This set of function calls from cognition uses three lists internal to the simulated ear: long-term targets, ignored targets, and short-term targets; and a flag to ignore interrupts. When something is heard, these functions synchronously or asynchronously return a heard object, which includes an ID that may be unique to the ear, spatial location or channel; onset time, duration, repetition-or-not; if language: speaker ID and sex, words spoken; modifiers of sound (e.g., walking *in grass*).

0. Asynchronously return the current sound.
1. Return the current sound.
2. Listen for a primary target in noise short term, given the noise or list of noises, return it asynchronously when found (add target to list). Return that the command has been received.
3. Listen for a secondary target as a secondary task. Return it asynchronously when found. (add target to list). Return that the command has been received.
4. Take interrupts from a list [passed the interrupt list or a delta to the existing list]. Return it asynchronously when found. Return that the command has been received.
5. Ignore interrupts on a list [passed the ignore list or a delta to the original list]. Return that the command has been received.
6. Long-term vigilance for a target sound [passed the list or a delta to the original list]. Return that the command has been received, and return it asynchronously when found.
7. Add- Remove- List items from the three lists. Return that the list has been modified.
8. Turn the head, either to an absolute or relative heading related to current heading. Return that the command has been received. Return that it has been done when done.

---

How well the simulated ear implements these functions defines what it means to have an ear. The functions might not always work, they might forget which list a found sound is on, for example, and might not be fully accurate on what is returned. The quality might be degraded to simulate a human, or the quality might be degraded because the auditory system used in the model is not even as good as a human's.

The auditory system also appears to run autonomously in many ways. Cognition will pass information to the auditory system, but the auditory system will be passing information continuously to cognition. Attention from cognition will decide when to use the auditory information.

The auditory system will include both asynchronous and synchronous communication with cognition. This may make the development and use of a CAS more complex. The human ear does more than just translate air pressure waves into "sound", but it also interprets and filters. A Computational Auditory System would have to replicate at least the basics of this.

## 5. Types of Sounds

We next describe the types of sounds that the ear can recognize. We draw our list from a longer report examining sound as used by Foley artists, in video games, and in the psychology literature. The constraints from the psychology literature are also taken up in Section 6.

The Appendix notes the types of sounds we found in our review. They include natural sounds, and sounds from humans and machines. This ontology appears to provide a way to organize all sounds, but it could be revised and is not currently complete. It might represent truly different kinds of sounds or it just might be a useful way to keep track of the diversity of sounds for system development.

## 6. Suggested Functionalities

Table 1's functions and the ontology of sounds (Appendix) to recognize can be examined and expanded in more detail. Table 2 notes capabilities and constraints on performance. They are numbered for current and later reference, and to help in later scorecards of how advanced each implemented CAS is. They are explained in more detail in a longer report.

One way to design and present updates about a model is to have a scorecard of the features of the data that the model should exhibit. A classic example is the list of regularities that Salthouse (1986) prepared for transcription typing. In his paper he noted 29 regularities. In a related model John compared her model's performance to these regularities (John, 1996). Thus, any single system at any point in time might not have all the capabilities, but might have some boxes filled in like a Yahtzee scorecard.

Section 1 of the scorecard in Table 2 notes some basic hardware (or equivalent) requirements. These might be realized in software if the sound comes in via an audio line. There may be further constraints, including that the microphones are appropriately positioned, such as 1 foot apart, and that a surrounding ear provides additional nuances to the sound. The volume sensitivity should be limited to 10 dB and 130 dB above background. It might be useful to start with the same frequency response as a human ear, but allow this to be modified to explore aging, hearing loss, and other scenarios.

Section 2 of Table 2 basic sound capabilities, notes that the CAS should be able to recognize the sounds in the Appendix, including natural (e.g., leaves rustling), machine made, and human-made sounds (e.g., dropped tool, walking). These sounds may be more different in the ontology for ease of manipulation than in how they are recognized. There are several aspects of sound that the CAS should be able to recognize. Not all sounds will be fully recognized. The spatial accuracy of recognition is not perfect and is affected by several factors (such as age and sound frequency) and is limited to approximately +/- 1-5° for a source in front of the listener (depending on frequency) and up to +/-20° for any on the side or behind (Boff & Lincoln, 1988, §2.812; Letowski & Letowski, 2011, p. 72).

*Table 2.* Capabilities and constraints on an artificial ear.

1. Basic hardware implementation
    a. Consist of two microphones to implement binaural processing.
    b. Sensitivity to frequency should be similar to a real ear (RE),
    c. Have a "front" and "sides". Many auditory effects are enhanced if the sound source is in front,
    d. Have the ability for different sensitivity curves to simulate damaged, younger/older ears.
2. Basic capabilities
    a. Recognize ambient sounds in the ontology in Table 2 and be able to identify the approximate distance and bearing of a sound using binaural processes.
    b. Be able to lock on to a target speaker or single auditory object and be able to track it in the presence of a variable noisy environment.
    c. Recognize sounds on primary target list.
    d. Recognize sounds on secondary target list.
    e. Primary and secondary list decays with time making probability of recognition decrease and time to recognize increase.
    f. Follow the target audio stream if ear moves spatially itself or if the target moves
    g. Frequency changes, Doppler effects. Detecting these and tracking despite these.
3. Speaker-related capabilities
    a. Identify voice types, genders, accents
    b. Identify speaker (this may be a cognitive effect, but some aspects may be in the ear and pre-processor, such as prosody, timbre, changes in volume, harmonics, vocabulary, etc.)
    c. Detecting effects on speech such as whispering, muttering, or speech impediments including lisping and talking with food in mouth
4. Recognition and cocktail party capabilities
    a. Hear listener's name or selected words in an audio stream
    b. Be able to focus on one audio stream, even in a relatively noisy background
    c. Be able to switch focus to a different audio stream based on a target of interest (such as an alarm, or its "name"). It should then be able to re-focus on the original audio stream.
5. Alarm-related capabilities and regularities
    a. ID each alarm sound for cognition to recognize it
    b. Recognize sound as new type or previously known type (might belong in cognition, but type and ID should support this).

These assumption and constraints might be modified to model a super-ear, an ear that can hear better than humans. But also see Schooler and Anderson (2005) that suggests that forgetting can help memory and cognition, and this is probably how people function. Having a super-ear might be distracting for typical tasks.

Section 3 of Table 2 covers speaker recognition and related effects. The system should be able to recognize a range of features of spoken language. Not every system and particularly not early systems will be able to recognize and report all these features. This may include the ability to recognize what language is being spoken and with what accent before or in addition to recognizing the words.

Section 4 of Table 2 is the ability to listen to a single audio stream in a complex environment. Here, there should be the ability to follow a single speaker and also to switch streams when one stream becomes more interesting.

Section 5 of Table 2 is the ability to recognize alarms. This set of regularities is related to generating interruptions to cognition. Some of the capabilities have to be added to the ear or cognitive architecture. An auditory pre-processor can do some of these tasks, or they can be done in cognition. The meaning of each alarm would be decided in cognition, but what is returned from the ear prior to this has to be unique enough for cognition to do its job. How significant the alarm is also has to be in cognition and may be based on context, but the volume and distance the sound is coming from appears to be the ear or pre-processor's job. A human factors book on alarms will be helpful as this area of capabilities is developed (Stanton, 1994).

The CAS should support cognition to the point where the ear and cognition represent how the alarm is raised to cognition in a way that helps recognize the alarm. Simply relaying the alarm is not any more useful than the alarm itself. The ear's alarm (content, volume, distance) representation has to support reasoning about context, meaning, possible action paths, and automatically opening up and displaying the relevant emergency procedures. These activities may all be supported by:

1. The ability to detune or not report an alarm stream (e.g., in the situation where there may be multiple ears paying attention in the same area like a hospital ward, or a submarine where each ear is focusing on a particular station). The ear should be able to recognize if the alarm is coming from its station and ignore other stations based on space or volume or another aspect of sound.
2. If there are many alarms at the same time, the CAS should be able to help cognition prioritize the alarms. Cognition or an auditory pre-processor might perform this task. If there are multiple alarms that are similar (such as all pumps have failed), the ear should consolidate them. Hollywell and Marshall (Stanton, 1994) concluded that operators can read alarms at a rate of 30 per min. if they had no other tasks, but preferred no more than 15 per min. Increased alarm rate does not decrease accuracy in assessing them, but does lead to missed alarms.
3. The ear can be programmed with knowledge of what alarms it will be listening for, but should be robust enough to identify an alarm that is not pre-programmed. New equipment can be installed, and current equipment could have a software update or other change that modified the sound or intensity of the alarm, which has happened with alarms.

## 7. Conclusion

We have provided an overview of the design of a computational audio system (an artificial ear) for cognitive architectures. Table 1 provides a list of functions an ear should provide to cognition. It also includes an outline of the data structures.

Appendix 1 provides a summary ontology of the types of sounds that an artificial ear could recognize and also example instances and sub-instances. We can see that we will need to be able to add sounds routinely to this list, and to code them and tie them to cognitive representations.

Table 2 provides a scorecard to represent how sophisticated and complete a given instantiation of a simulated ear is. Together, they provide a design for a computational audio scene analysis for use by cognitive architectures and agents.

The list of capabilities and constraints are incomplete. As we develop this initial list we have found more challenging items that we have not yet added, including habituation, other types of learning, a phonological loop, and audio priming.

We have started work on implementing an artificial ear for use by cognitive architectures. We have created a capability to duplicate the cocktail party effect by using cluster analysis on cochleagram (audio spectrograms). Reports are available on its technical implementation (Daley, Bonacci, Gever, Diaz, & Bolkhovsky, 2021) and how it is being integrated with the Arcadia architecture (Gever et al., 2020). In this integrated model the ear will be used to turn a head towards a sound and look at it.

This work suggests several insights. The greatest is to the extent that there is active vision (Findlay & Gilchrist, 2003), which there is, there will be active hearing as well. Thus, there will be interactions between the ear, cognition, and action; they will work together where cognition receives information from the ear, and in turn, passes to the ear sounds to listen for, or to modify the hearing apparatus (e.g., turn the head) to hear better. Where this information processing occurs might vary depending on the implementation of the CAS and the architecture. As we create this model ear, we will need example applications to show this interaction.

## Acknowledgments

Support for this project was provided as work unit F1103–Office of Naval Research, Code 34. Lia Bonacci provided very used comments. Steve Croker, David Gever, and the reviewers provided useful comments. Sue Van Vactor provided a useful proofreading, twice. The views expressed in this article reflect the results of research conducted by the authors and do not necessarily reflect the official policy or position of the Department of the Navy, Department of Defense, nor the United States Government. The authors are federal and contracted employees of the United States government. This work was prepared as a part of official duties. Title 17 U.S.C. 105 provides that "copyright protection under this title is not available for any work of the United States Government." Title 17 U.S.C. 101 defines a U.S. Government work as work prepared by a military service member or employee of the U.S. Government as part of that person's official duties.

# Appendix 1. Ontology of sounds (partial).

| Type | Instances | Features for feature ontology |
|---|---|---|
| **Natural** | | |
| | Mammals | |
| |   Dog | Type of sound (bark, growl, etc.) |
| |   Horse | |
| | Birds | Type (e.g., woodpecker), Action: flapping wings, call |
| | Insects | Type |
| | Leaves/wind | |
| | Fluid sounds | Dripping, droplets, flushing, etc. |
| | Water | |
| | | Snow, rain, thunder |
| | Fire | |
| | Rocks | |
| **Human and Human Made** | | |
| | Speech | speaker, volume, gender, emotional tone, words |
| | Singing | |
| | Whistling | |
| | Retching, spitting | |
| | Gasping, yelling, whimpering, moaning, | |
| | Breathing, blowing nose, sneezing, coughing | |
| | Footsteps | Terrain (e.g., puddle, firm ground) |
| | Getting hit | e.g., with a stick |
| | Heartbeat | |
| **Mechanically-made by humans or machines** | | |
| | Alarms & Warnings | |
| | Tapping | Pace, material tapped |
| | Click, dongs, dings | |
| | Creaking | Chair, fence |
| | Dragging | |
| | Gunfire | Distance |
| | | Caliber |
| | | Rate of Fire |
| | | Direction |
| | Grenade | |
| | Gear rustling | |
| | Magazine changes/other clicking sounds | |
| | Trucks/vehicles | Speed |
| | | Direction |
| | | Distance |
| | | Horn |
| | | Type of Vehicle |
| | Explosions | Size |
| | | Distance |
| | | Direction |
| **Miscellaneous Sounds** | | |
| | "Sha" (unknown word) | |
| | | Standard features |
| | Music | many features |
| | Unknown | |